\documentclass[aps,prd,reprint,amsmath,amssymb,superscriptaddress,showkeys,noeprint]{revtex4-1}
\usepackage{graphicx}
\usepackage{bm}
\usepackage{units}
\usepackage{hyperref}
\usepackage[mathlines]{lineno}
\usepackage{hepnames}

\begin{document}

\newcommand{\QGSJETIIc}{\textsc{QGSJetII.03}}
\newcommand{\QGSJETIId}{\textsc{QGSJetII.04}}
\newcommand{\EPOSLHC}{\textsc{epos~lhc}}
\newcommand{\EPOSThree}{\textsc{epos3}}
\newcommand{\PYTHIA}{\textsc{pythia}}
\newcommand{\abs}[1]{\left|{#1}\right|}
\newcommand{\avg}[1]{\langle{#1}\rangle}
\newcommand{\Esub}[1]{E_{#1}}
\newcommand{\EsubText}[1]{\Esub{\mathrm{{#1}}}}
\newcommand{\density}[1]{\mathrm{d}{#1}/\mathrm{d}\eta}
\newcommand{\densityN}[1]{\mathrm{d}{#1}/\mathrm{d}N_{\mathrm{ch}}}
\newcommand{\lna}{\avg{\ln\!A}}
\newcommand{\xmax}{X_\text{max}}
\newcommand{\nch}{N_{\rm ch}}
\newcommand{\nmu}{N_\mu}
\newcommand{\lnn}{\ln\!\nmu}
\newcommand{\fcore}{\omega_\text{core}}

\title{Core-corona effect in hadron collisions and muon production in air showers}

\author{Sebastian Baur}
\email{sebastian.baur@ulb.ac.be}
\affiliation{%
Universit{\'e} Libre de Bruxelles, IIHE -- CP230, B-1050 Brussels, Belgium
}%
\author{Hans Dembinski}
\affiliation{
Max Planck Institute for Nuclear Physics, Heidelberg, Germany
}%
\author{Matias Perlin}
\affiliation{%
Institute for Nuclear Physics,
Karlsruhe Institute of Technology, Karlsruhe, Germany
}%
\affiliation{%
Instituto de Tecnolog\'ias en Detecci\'on y Astropart\'iculas (CNEA, CONICET, UNSAM), Buenos Aires, Argentina
}%
\affiliation{%
Departamento de F\'isica, FCEN, Universidad de Buenos Aires, Buenos Aires, Argentina
}%
\author{Tanguy Pierog}
\affiliation{%
Institute for Nuclear Physics,
Karlsruhe Institute of Technology, Karlsruhe, Germany
}%
\author{Ralf Ulrich}
\affiliation{%
Institute for Nuclear Physics,
Karlsruhe Institute of Technology, Karlsruhe, Germany
}%
\author{Klaus Werner}
\affiliation{
SUBATECH, University of Nantes -- IN2P3/CNRS -- IMT Atlantique, Nantes, France 
}%

\date{\today}

\begin{abstract}
It is very well known that the fraction of energy in a hadron
collision going into electromagnetic particles (electrons and
photons, including those from decays) has a large impact on
the number of muons produced in air shower cascades. Recent
measurements at the LHC confirm features that can be linked to a
mixture of different underlying particle production mechanisms such as
a collective statistical hadronization (core) in addition to the
expected string fragmentation (corona). Since the two mechanisms have
a different electromagnetic energy fraction, we present a possible connection between
statistical hadronization in hadron collisions and muon production in
air showers.  Using a novel approach, we demonstrate that the
core-corona effect as observed at the LHC could be part of the solution for
the lack of muon production in simulations of high energy cosmic rays.
To probe this hypothesis, we study hadronization in high
energy hadron collisions using calorimetric information over a large
range of pseudorapidity in combination with the multiplicity of
central tracks. As an experimental observable, we propose the
production of energy in electromagnetic particles  versus hadrons, as a function of
pseudorapidity and central charged particle multiplicity.
\end{abstract}

\maketitle

\section{Introduction}
Cosmic ray particles reach Earth from galactic and extragalactic
sources with enormous energies and produce huge particle cascades in
the atmosphere. The resulting extensive air showers are measured with
the aim to unveil the astrophysical nature and origin of high energy
cosmic rays. The Pierre Auger
Observatory~\cite{ThePierreAuger:2015rma,Abraham:2009pm} and the
Telescope Array~\cite{Tokuno:2012mi,Tokuno:2012mi} are the largest
contemporary experiments targeting the most energetic cosmic rays with
energies beyond $10^{18}$\,eV.

Of particular interest is the cosmic ray mass composition,
which is expected to carry a unique imprint of the physics at the sources.
The mass composition as a function of the cosmic ray energy $E_0$ is inferred from air shower
observables, of which the most important ones are the depth of the
shower maximum $\xmax$ and the number of muons
$\nmu$~\cite{Kampert:2012mx}. The depth $\xmax$ is the integrated
matter density column that a shower traversed until the maximum number
of charged particles in the shower is reached. The number of muons is
obtained by counting muons when the shower arrives at
ground. Experimentally the muon counting is limited to a radial range
around the shower axis as well as to a minimal energy of muons. 

To infer the cosmic ray mass composition from these observables,
accurate predictions from air shower simulations are needed for cosmic
rays with various primary masses. However, the Pierre Auger
Observatory~\cite{Aab:2014pza,Aab:2016hkv} and the Telescope
Array~\cite{Abbasi:2018fkz} observed that the measured number of muons
in air showers drastically exceeds expectations from model predictions
at shower energies around and above $10^{19}\,$eV.
A recent summary of muon measurements~\cite{Dembinski:2019uta} shows that
a consistent muon excess is seen by the majority of cosmic ray
experiments over a very wide energy range. The discrepancy between results based on $\xmax$ and $\nmu$ is
currently preventing an unambiguous interpretation of air shower data 
in terms of mass composition.

The amount of energy ending up in
electromagnetic particles in hadron collisions
\begin{equation}
  \label{eq:r}
  R=\frac{\EsubText{em}}{\EsubText{had}}, 
\end{equation}
where $\EsubText{em}$ is the summed energy over all $\gamma$ (mostly from
$\pi^0$ decay) as well as $e^{\pm}$,  and $\EsubText{had}$ the summed energy of all
hadrons, is one of the crucial parameters driving muon production in
extensive air showers~\cite{Ulrich:2010rg,Cazon:2018gww,Cazon:2018bvs}. It is closely
related to the way an excited partonic system hadronizes. In hadronic
interaction models used to simulate air showers, the hadronization is
mainly done using a string fragmentation model which was successfully
developed to describe the hadron production in $e^+$-$e^-$ collisions,
and low energy proton-proton collisions. In systems with higher energy
densities, such as heavy ion collisions, 
a statistical hadronization of a fluid is expected
where the production of heavy particles
is favored, thus, reducing the fraction of $\pi^0$ compared to other
types of particles.  In the early 2000s ``collective effects'' have
been observed in heavy ion collisions (often referred to as
\emph{large} systems) at
RHIC~\cite{Adams:2005dq,Adcox:2004mh,Arsene:2004fa,Back:2004je}. Similar
effects have been
predicted~\cite{Werner:2010ss,Bozek:2010pb,dEnterria:2010xip,Prasad:2009bx,Ortona:2009yc,Cunqueiro:2008uu}
for proton-proton collisions (aka \emph{small} systems) and were
eventually discovered at the LHC~\cite{Khachatryan:2010gv} (see
Refs.~\cite{Dusling:2015gta,Loizides:2016tew} for detailed
reviews).

While a fluid-like behavior (referred to as collective effects in the following) is confirmed in both large and small systems,
their origin is still unclear. In large systems the
existence of a quark-gluon-plasma (QGP) is commonly assumed as a phase
of parton matter where confinement is no longer
required~\cite{Shuryak:1980tp,Stoecker:1986ci,Kolb:2003dz}.  This QGP
will evolve according to the laws of hydrodynamics and eventually
decay statistically.  There are various expected
consequences of such a scenario, such as long-range two-particle
correlations, the so-called ``ridge''
phenomenon~\cite{Abelev:2009af,Khachatryan:2010gv}, jet
quenching~\cite{Aad:2010bu,Chatrchyan:2011sx}, or enhanced production
of strange hadrons~\cite{ALICE:2017jyt}.
It was initially a surprise when such effects were also discovered in
small systems. While it was argued that also in central collisions of
small systems the energy densities may be high enough to allow for the
formation of a QGP~\cite{Werner:2010ss}, other recent studies have
shown that collective effects can be achieved by alternative
mechanisms such as microscopic effects in string
fragmentation~\cite{Bierlich:2017vhg} or QCD
interference~\cite{Blok:2017pui}.
The possibility  of collective effects in smaller systems
opens the door to study the impact of a different hadronization scheme
in high energy interactions also within air showers.  Air shower
cascades are driven by collisions of hadrons and light nuclei at
ultra-high energies. We show that statistical hadronization in
collisions of hadrons and nuclei can play a so far underestimated
importance in the understanding of muon production in air
showers~\cite{Pierog:2019opp,Anchordoqui:2019laz}.

The underlying mechanism responsible for the production of these
effects is expected to produce characteristic observables in the final
state of hadron collisions. We demonstrate how statistical hadronization
affects the energy fraction contained in electromagnetic versus hadronic
particles, $R$,  and show how this has important possible
implications for the muon production in cosmic ray air showers.

We further propose detailed measurements of $R$ as a novel opportunity to study collective
hadronization in small systems at the LHC. This may lead to a better
understanding of the underlying nature of statistical hadronization
since different theoretical approaches lead to predictions that
may be distinguished based on measurements.
In addition those measurements are able to constrain models for air shower simulations.

\section{The muon problem and the $R$ observable}
\label{sec:R}

The dominant mechanism for the production of muons in air showers is
via the decay of light charged mesons. The vast majority of mesons are
produced at the end of the hadron cascade after typically five to ten
generations of hadronic interactions (depending on the energy and
zenith angle of the cosmic ray).
The energy carried by neutral pions, however, is directly fed to the
electromagnetic shower component and is not available for further
production of more mesons and subsequently muons. The energy carried
by hadrons that are not neutral pions is, on the other hand, able to produce
more hadrons and ultimately muons in following interactions and
decays. Using a simple Heitler type toy-model~\cite{Pierog:2006qv}
based on~\cite{Matthews:2005sd}, the \emph{neutral pion fraction} $c
= N_\mathrm{\pi^0}/N_\mathrm{mult}$, defined as the number of neutral
pions $N_\mathrm{\pi^0}$ divided by the total number of final-state particles $N_\mathrm{mult}$ in a collision, was found
to have a strong impact on the muon number and in particular on the
slope of the energy dependence of the muon production. Indeed in this
model we get
\begin{equation}
  N_{\mu}=\left( \frac{\Esub{0}}{\EsubText{dec}} \right)^\beta \;\;\;\text{with}\;\;\;\;
  \beta={ 1+\frac{{\rm ln}(1-c)}{{\rm ln}N_{\rm mult} }}, 
\label{eq:kw1}
\end{equation}
where $\Esub{0}$ is the energy of the primary cosmic ray particle and $\EsubText{dec}$ is
the typical energy at which mesons decay in the cascade. 
So the muon number $N_{\mu}$ increases strongly with decreasing $c$, which is understandable since more hadrons
is available to produce muons. 
A second quantity with a strong impact on the muon number was identified
to be the hadron multiplicity $N_\text{mult}$. 

The value of $c$ is very important for the muon
production. Unfortunately, it is 
difficult to measure both
$N_\mathrm{\pi^0}$ and $N_\mathrm{mult}$ experimentally (for example at the LHC) since neutral
particles cannot be easily counted individually. In general, 
secondary particle identification is unavailable at large
pseudorapidities $\eta$ where the energy flow is large enough to
become relevant for the air shower development.  Hence, we propose a new
observable which is sensitive to properties of the hadronization 
and which can be directly related to $c$: the ratio of
the electromagnetic to the hadronic energy density $R$ given by
\begin{equation}
  \label{eqn:ratio}
R(\eta) = \frac{ \avg{ \density{\EsubText{em}}} }{\avg{\density{\EsubText{had}}}} \,\text{.}
\end{equation}
Here the energy densities $\avg{\density{E}}$ are obtained by summing
the energy of all final-state particles except for neutrinos in bins
of $\eta$ and averaging over a large number of collisions. 

The neutral pion fraction $c$ can be easily related to the energy
ratio $R$, since both are very similar kinematic aspects of final
state distributions. If all particles have
the same energy such as in the generalized Heitler model, then we have
simply $R=c/(1-c)$. But $R$ is experimentally much easier to measure,
since, using a calorimeter, the signals deposited by electromagnetic
particles and by hadrons are characteristically different.  We compute
a detailed conversion between $R$ and $c$ using standalone \EPOSLHC{}~\cite{Pierog:2013ria}
simulations of fixed
energy proton-proton collisions at various center-of-mass energies, and
found that for the relevant parameter range, a change of $R$ by $\Delta R$
affects $c$ by $\Delta
c\approx0.8\cdot\Delta R$, where $R$ is computed by integrating
eq.~(\ref{eqn:ratio}) over all $\eta$.
In section~\ref{sec:LHC}, we will study $R$ for different models 
as a function of $\eta$, and at fixed $\eta$ as a function of the
charged particle density at central pseudorapidity
$\left.\density{N_{\mathrm{ch}}}\right|_{\eta=0}$, which is determined
as the average multiplicity within $\abs{\eta}<0.5$.

The influence of various effective parameters $q$
 in interaction models (like $R$, $c$, or $N_\textrm{mult}$) on
the main air shower observables 
was investigated in a previous
study~\cite{Ulrich:2010rg} in which the
behavior of hadronic interaction models in air shower simulations was
modified in an energy-dependent way during full air
shower cascade simulations within CONEX~\cite{Bergmann:2006yz}.

The effective quantity $q$ of the hadronic event generators inside the
air shower cascade simulation is changed in an energy dependent way
\begin{equation}
  \label{eq:q}
 q(\EsubText{lab}) \to q(\EsubText{lab}) \times \bigl( 1 + f_q  F(\EsubText{lab};\EsubText{th},\EsubText{scale})\bigr)  
\end{equation}
using the modification scale $f_q$, and the energy-dependent factor
\begin{equation}
  \label{eq:F}
  F(\EsubText{lab};\EsubText{th},\EsubText{scale})=\frac{\log_{10}(\EsubText{lab}/\EsubText{th})}{\log_{10}(\EsubText{scale}/\EsubText{th})} \; \text{for} \; \EsubText{lab}>\EsubText{th},
\end{equation}
representing the assumption that models are well constraint by
accelerator data at lower energies (below $\EsubText{th}$), where
$F(\EsubText{lab})=0$, while they become logarithmically unconstrained
going to higher energies. The parameter $\EsubText{scale}$ is the \emph{reference} energy scale. We will use $\EsubText{scale}=E^{\rm
  CR}_{\rm LHC} \simeq s_{\rm LHC}/(2m_{\rm p})\approx\unit[90]{PeV}$, using
an LHC center-of-mass energy of 13\,TeV. Typical threshold values are
$\EsubText{th} \simeq s_{\rm Tevatron}/(2m_{\rm p})\approx1\,$PeV, using the
center-of-mass energy of the Tevatron accelerator. However, in
particular particle production, in the important forward phase space, may be
largely unconstrained by both Tevatron and LHC data, allowing much
lower values of $\EsubText{th}$ to be explored. It is a key point
of the application of eq.~(\ref{eq:F}) inside CONEX that a
significant fraction of the air shower cascade is consistently
modified during the simulations.

\begin{figure}
  \includegraphics[width=\columnwidth,trim=10 10 25 0]{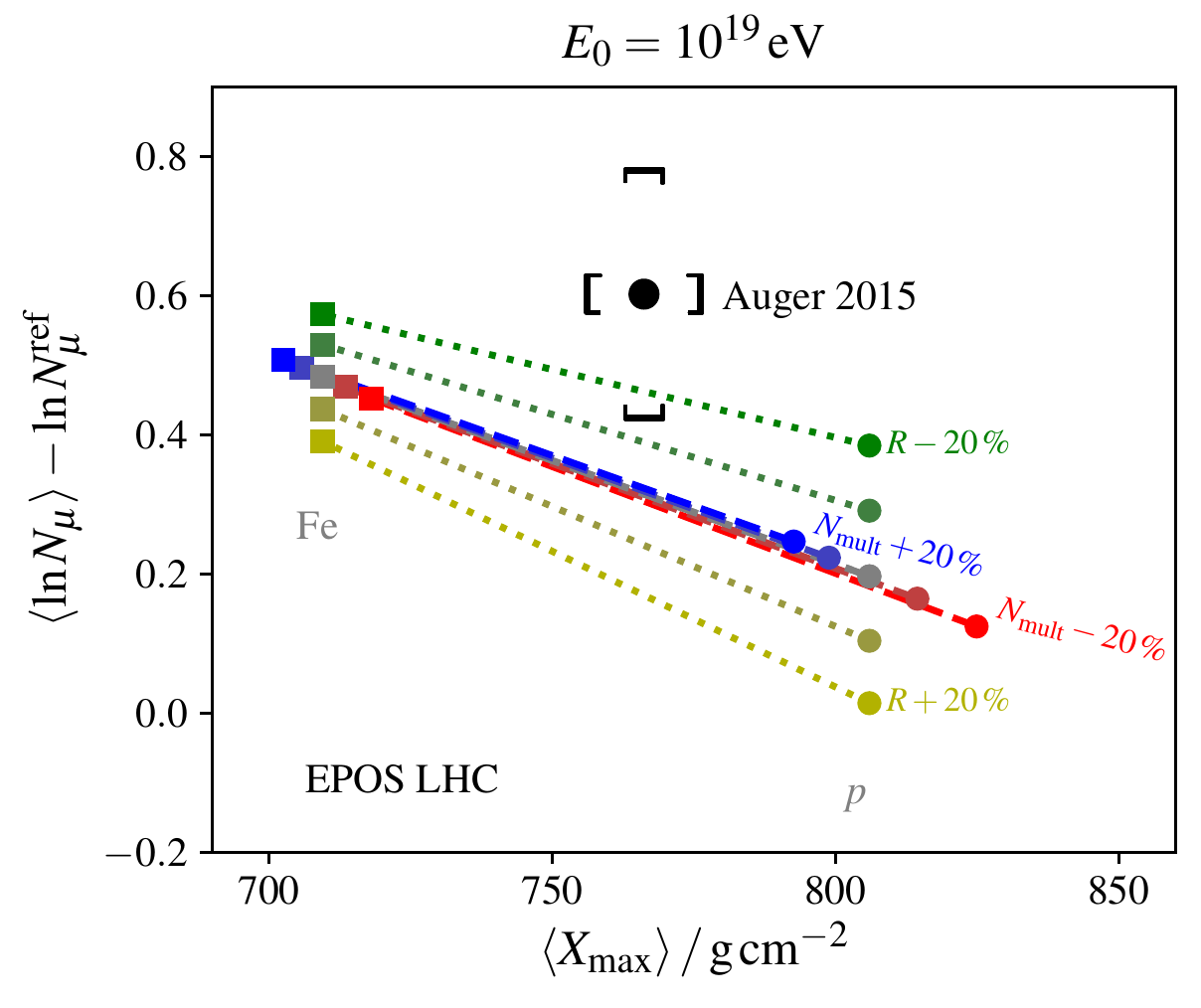}
  \caption{Impact of the modification scales $f_q$ (in \%)
  of the hadron multiplicity $N_\text{mult}$ (dashed lines) and the
  energy ratio $R$ (dotted lines) in collisions at the LHC energy of
  $\sqrt{s}=$13\,TeV on \EPOSLHC{} predictions of the air shower
  observables $\xmax$ and $\lnn$ in $10^{19}$\,eV air
  showers. The datum is from the Pierre Auger
  Observatory~\cite{Aab:2014pza}. The model lines represent all values
  that can be obtained for any mixture of cosmic nuclei from proton
  (bottom right) to iron (top left). }
  \label{fig:air_shower_muons}
\end{figure}

We apply eqs.~(\ref{eq:q},\ref{eq:F}) to explore the correlated impact of  $q=R$ and $q=N_\text{mult}$  on $\xmax$ and $\lnn$ in full air shower simulations. The resulting correlated effect is shown in
Fig.~\ref{fig:air_shower_muons} as demonstrated for air showers at
$\Esub{0}=10^{19}$\,eV using \EPOSLHC{} in CONEX.  Lines in this
figure show all possible resulting mean values of $\xmax$ and $\lnn$
for any mass composition of cosmic rays between pure proton (bottom
right end of lines) and pure iron (top left end of lines).
The resulting values of $\xmax$ and $\lnn$ are located on a straight
line because the mean values for both are linear functions of the
mean-logarithmic mass of cosmic
rays~\cite{Abreu:2013env,Dembinski:2017kpa} given a fixed air shower
energy. The line-shape is universal, but its location, and to a lesser
degree the slope and length, depend on the hadronic interaction model.
Current hadronic interaction
models predict lines, which are too low compared to experimental data
from air showers, as indicated by the vertical gap between the
representative data point from the Pierre Auger
Observatory~\cite{Aab:2014pza} and the \EPOSLHC{} line. This
discrepancy is the expression of the muon problem outlined above.

When $N_\text{mult}$ is modified  the simulated line shifts along 
itself: the multiplicity has a correlated effect on $\xmax$ and $\lnn$
that cannot close the gap to the data.  However, modifications of $R$ mainly
affect the muon number and leave $\xmax$ unchanged, creating vertical
shifts and tilts of the line in the plot.  Thus, within the
assumptions outlined here, we find that a decrease of $R$ by $f_q=-15\%$ at
the LHC energy of $\sqrt{s}=13$\,TeV would be sufficient to make the
simulations compatible with the air shower data at $10^{19}$\,eV.
These results have been cross-checked with alternate interaction
models in the air shower simulations. There is a very good qualitative
agreement in all cases.

Furthermore, in Ref.~\cite{Dembinski:2019uta} it was established
that the muon discrepancy
in simulations increases smoothly with energy. Thus, the slope of the
energy dependence introduced in eq.~(\ref{eq:kw1}) is also affected, pointing to a too small value of $\beta$.
This may be related to a too large $\pi^0$ production. We explore this energy dependence in more detail in
the next section.

\section{Core-corona effect and muon problem }
\label{sec:cr}

The discussion in the previous section suggests that a change of $R$ (or
$c$, which is equivalent) is a potential way to
reduce the discrepancy between measurements and air shower
simulations. Nevertheless, $R$ is quite well constrained by theory as well as laboratory
measurements and, thus, can not be changed entirely arbitrarily as studied in the previous section~\ref{sec:R}.
In a naive model like Ref.~\cite{Matthews:2005sd} where
only pions are considered as secondary particles, $R=0.5$. In a
more realistic approach based on string fragmentation we have 
$R\approx0.41$. 
But as shown
in Ref.~\cite{ALICE:2017jyt}, particle ratios such as $K/\pi$, $p/\pi$ or
$\Lambda/\pi$ change with increasing secondary particle density, saturating to the value
given by a thermal/statistical model with a freezeout temperature of
156.5\,MeV~\cite{Andronic:2016nof} yielding 
$R\approx0.34$.  Such a behavior can be explained in terms of a
core-corona picture \cite{Werner:2018yad}. This approach has been used in
the framework of realistic simulations \cite{Werner:2007bf}, but also
in simple model calculations~\cite{Manninen:2008mg,Becattini:2008yn,Aichelin:2010ed,Aichelin:2008mi}.
The basic idea is that some
fraction of the volume of an event (or even a fraction of events) behaves as a quark gluon plasma and decays
according to statistical hadronization (core), whereas the other part
produces particles via string fragmentation (corona).  The particle
yield $N_i$ for particle species $i$ is then a sum of two
contributions
\begin{equation}
  \label{eq:coco}
   N_i = \fcore \, N_i^{\mathrm{core}} + (1-\fcore) \, N_i^{\mathrm{corona}},
\end{equation}
where $N_i^{\mathrm{core}}$ represents statistical (grand canonical)
particle production, and $N_i^{\mathrm{corona}}$ is the yield from
string decay. Crucial is the core weight $\fcore$. In order to
explain LHC data~\cite{ALICE:2017jyt} the weight
$\fcore$ needs to increase monotonically with the multiplicity,
starting from zero for low multiplicity $\Pproton{}\Pproton{}$ scattering, up to 0.5 or
more for very high multiplicity $\Pproton{}\Pproton{}$, reaching unity for central
heavy ion collisions (PbPb).

In the following, we are going to employ a straightforward core-corona
approach, based on eq.~(\ref{eq:coco}), for any hadronic interaction
model in CONEX air shower simulations. The particle yield from the
chosen interaction model is by definition considered to be the corona
yield, whereas we use the standard statistical hadronization (also
referred to as resonance gas) for the core part. So $\fcore=0$
would be the ``normal" simulation with the default interaction
model. Choosing $\fcore>0$ amounts to mixing the yields from
the interaction model according to the core-corona superposition shown in
eq.~(\ref{eq:coco}).  The core will certainly help concerning the
``muon problem", because statistical hadronization produces more heavy
particles and less pions compared to string fragmentation, and
therefore $R$ is smaller~\cite{Pierog:2019opp,Anchordoqui:2019laz}.

Technically, we directly modify individual particle ratios of the
secondary particle spectra ${\rm d}N_i/{\rm d}\Esub{j}$, for particle
species $i$ and energy bins ${\rm d}\Esub{j}$, of hadronic interactions
with air nuclei used by CONEX for numerical air shower 
simulations based on cascade equations.  Knowing the initial ratios
$\pi^0/\pi^\pm, p/\pi^\pm, K^\pm/\pi^\pm, p/n, K^0/K^\pm$ (taking into
account strange baryon decays) from a corona type model and the value
of the same ratios from the core model, we compute new spectra in
which the particle yields include both, core and corona according to
$\fcore$. Since the hadronization mechanism can affect only newly
produced particles the properties of the leading particle should be
preserved. To achieve that, the new particle yields are computed for
all secondaries, but excluding the one corresponding to the respective
projectile type, i.e.~protons in proton-air, kaons in kaon-air
interactions, and so on.  The yield of the projectile-type particles is
determined subsequently by exploiting energy conservation in all
energy bins ${\rm d}\Esub{j}$ summed over all secondary particle
species $i$: the sum $\sum_i \Esub{j} {\rm d}N_i/{\rm d}\Esub{j}$ must
be conserved.  Since at high $x_{\rm F}=\EsubText{j}/\EsubText{lab}$
only the projectile-type particles will have ${\rm d}N_i/{\rm
  d}\Esub{j}$ significantly different from zero (aka leading-particle
effect), the resulting modified leading-particle type spectra at high $x_{\rm F}$ follow
the original distribution, and are only affected by the scaling procedure at lower values of $x_{\rm F}$.
Together, this assures that energy conservation as well as the total
multiplicity are not affected, but only the particle ratios. More
details will be given in a future publication.

\begin{figure}
  \includegraphics[width=\columnwidth]{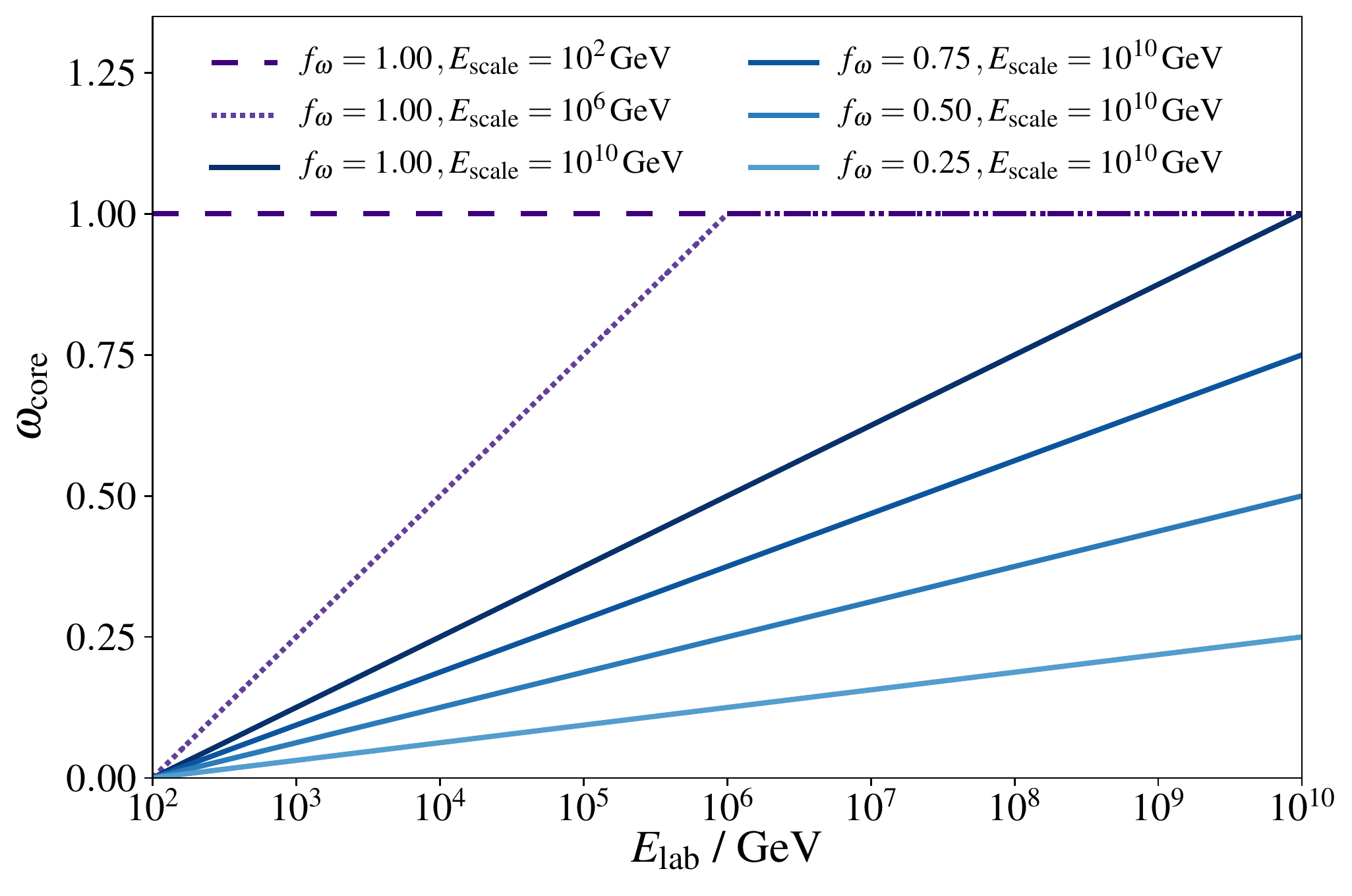}
  \caption{Different energy evolutions probed for $\fcore$. The solid
    lines represent changing the scale $f_\omega$ of the effect, while the
    dashed lines also indicate the effect of changing
    $\EsubText{scale}$. }
  \label{fig:fcore}
\end{figure}

\begin{figure*}
  \includegraphics[width=.8\columnwidth]{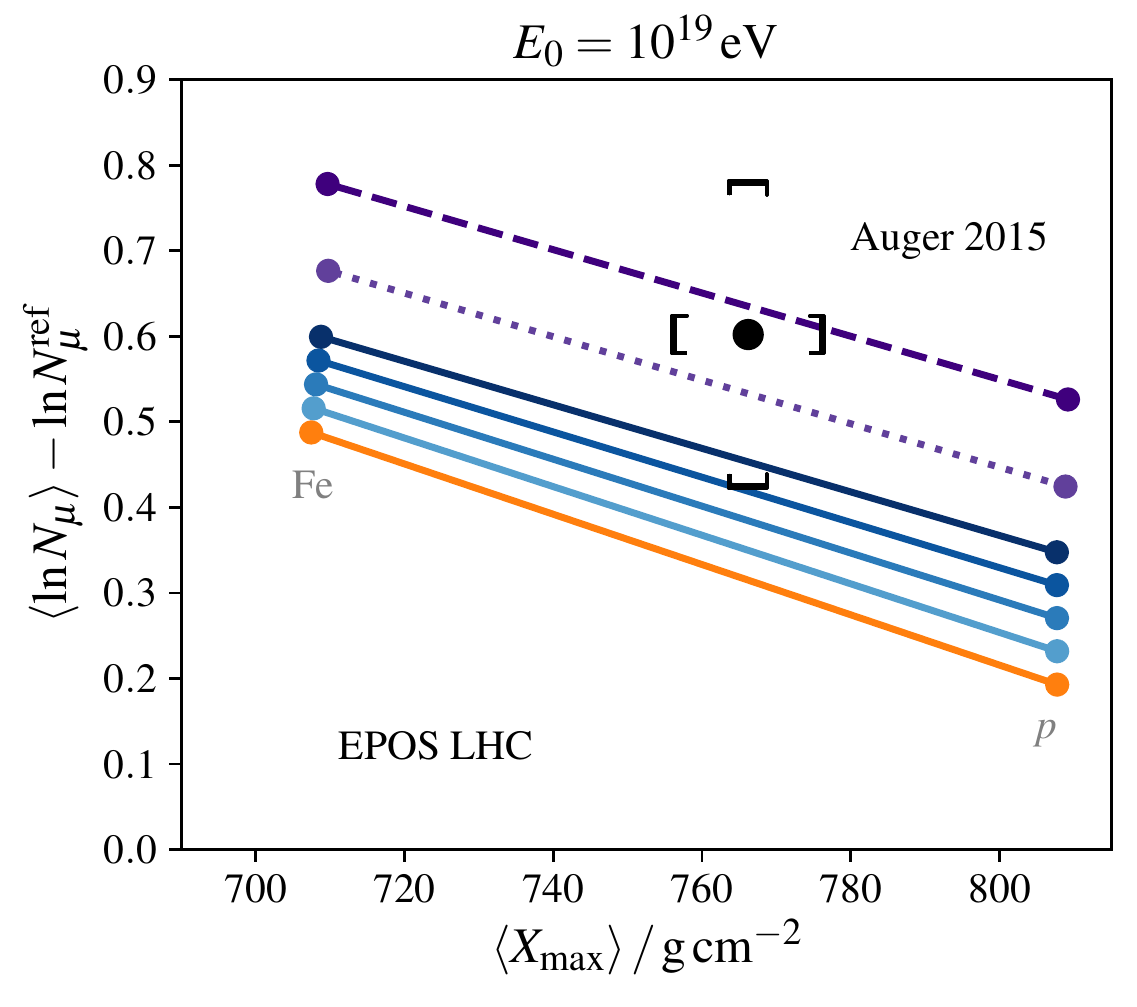}
  \includegraphics[width=1.26\columnwidth]{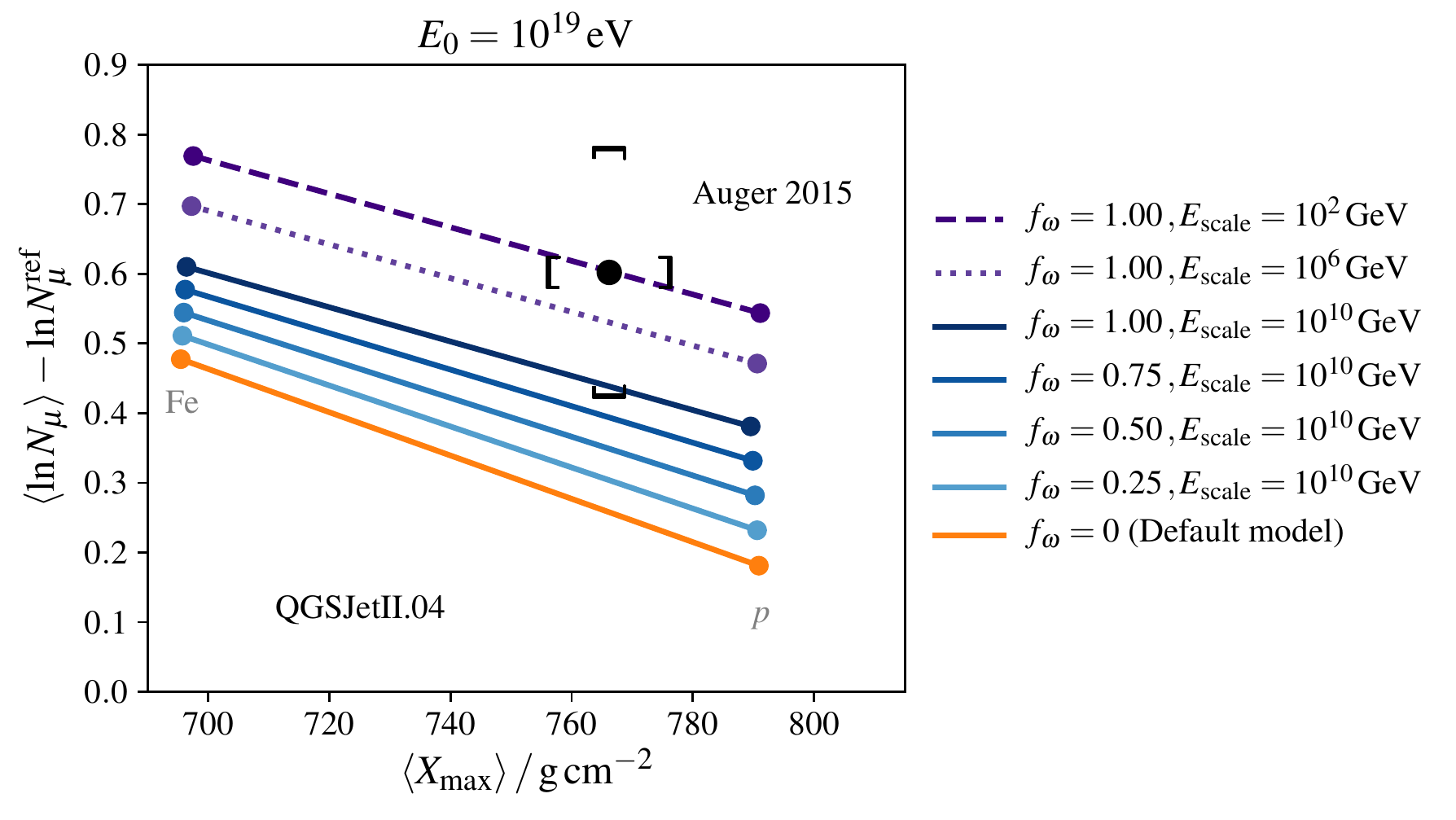}
  \caption{Comparison of different core-corona mixing scenarios, as
    described in the text, on air shower simulations at $10^{19}\,$eV
    using \EPOSLHC{} (left) and \QGSJETIId{} (right) in the
    $\xmax$-$\lnn$ plane.  The solid lines represent changing the
    scale $f_\omega$, while the dashed lines also
    indicate the effect of changing $\EsubText{scale}$. The \emph{default} model
    corresponds to the corona-only simulations. The
    datum is from the Pierre Auger Observatory~\cite{Aab:2014pza}. Each 
    model line represents all values that can be obtained for any
    mixture of cosmic nuclei from proton (bottom right) to iron (top
    left). }
  \label{fig:lnNmu_xmax}
\end{figure*}

We expect the core weight $\fcore$ to increase with energy in a
logarithmic way. Thus, we use
\begin{equation}
  \fcore(\EsubText{lab})=f_\omega \, F(\EsubText{lab};\EsubText{th},\EsubText{scale})
\end{equation}
to model this (in analogy to eq.~(\ref{eq:q})), starting
already at fixed-target energies, $\EsubText{th}=100\,$GeV.  Different
energy dependencies are explored by changing $\EsubText{scale}$ from
$100\,$GeV (corresponding to a step function), to $10^6\,$GeV, and
$10^{10}\,$GeV. The $f_{\omega}$ scale is varied from 0.25, 0.5, 0.75 to
1.0; in addition we enforce
$F(\EsubText{lab};\EsubText{th},\EsubText{scale})\overset{!}{=}1$ for all
$\EsubText{lab}\ge\EsubText{scale}$. This yields the $\fcore$ energy
dependencies as depicted in Fig.~\ref{fig:fcore}.
All these scenarios have been used to simulate full air showers with CONEX, 
using cascade equations from the first interaction 
to the ground, for proton and iron primary particles at $\Esub{0}=10^{19}\,$eV.
In Fig.~\ref{fig:lnNmu_xmax} the results
are shown in the $\xmax$-$\lnn$ plane for two models \EPOSLHC{} (left)
and \QGSJETIId{}~\cite{Ostapchenko:2006vr,Ostapchenko:2010gt} (right).
These examples illustrate
that it is well possible with modified hadronization in air
shower cascades to describe the data of the Pierre Auger
Observatory.  As expected, more core-like contributions are needed
compared to what is currently provided by the models.  This means, QGP-like
effects also in light colliding systems and starting in central
collisions at much lower center-of-mass energies may play a decisive
role. 

Furthermore, from eq.~(\ref{eq:kw1}) also a different energy evolution
of the muon production follows. To study the effect of our core-corona model on
the muon production as a function of the energy, we can compare the
different scenarios with the compilation of data presented
in Ref.~\cite{Dembinski:2019uta} using the renormalized factor 
\begin{equation}
    z = \frac{\langle \ln N_\mu \rangle - \langle \ln N_\mu
      \rangle_{\mathrm{p}}}{\langle \ln N_\mu
      \rangle_{\mathrm{Fe}}-\langle \ln N_\mu
      \rangle_{\mathrm{p}}},
    \label{eq:zfact}
\end{equation}
with $N_\mu$ being any muon related experimental observable and
$\langle \ln N_\mu \rangle_{\mathrm{p}}$ and $\langle \ln N_\mu
\rangle_{\mathrm{Fe}}$ being the average of the logarithm of the same
observable simulated with proton and iron primaries respectively for a
given reference hadronic interaction model.
This allows a direct comparison between different experiments for
various types of muon observables.

Considering the energy dependence of $z$, there is an implicit dependence on the cosmic-ray mass $A$, since $\langle\mathrm{ln}\,A\rangle$ varies with energy. However, as expected from the Heitler model formula, and even more importantly, verified via  explicit simulations, $z$ and $\langle \mathrm{ln}\,A \rangle$ are related as $z=a+b\langle\mathrm{ln}\,A\rangle$, and from $z(\textrm{pure Fe})=1$ and $z(\textrm{pure p})=0$ we simply get $a=0$ and $b=1/\mathrm{ln}56$. This is very useful, since it means that the $A$-dependence of $z$  (called $z_{\textrm{mass}}$) is given as
\begin{equation}
  z_{\mathrm{mass}}=\frac{\langle \mathrm{ln}\,A \rangle}{\mathrm{ln}\,56},
    \label{eq:zmass}
\end{equation}
and the expectation of $\Delta z=z-z_{\mathrm{mass}}$ is zero for the case of full consistency between all
experimental observables and the simulations based on a valid reference model. 
This means, plotting $\Delta z$ for experimental data, we should get zero if the reference model were perfect, whereas $\Delta z>0$  implies a muon deficit in the simulations. In this way we can visualize the energy dependence of the muon excess, corrected for mass dependencies. More details and references are given in Ref.~\cite{Dembinski:2019uta}.  

\begin{figure*}
  \includegraphics[width=\textwidth]{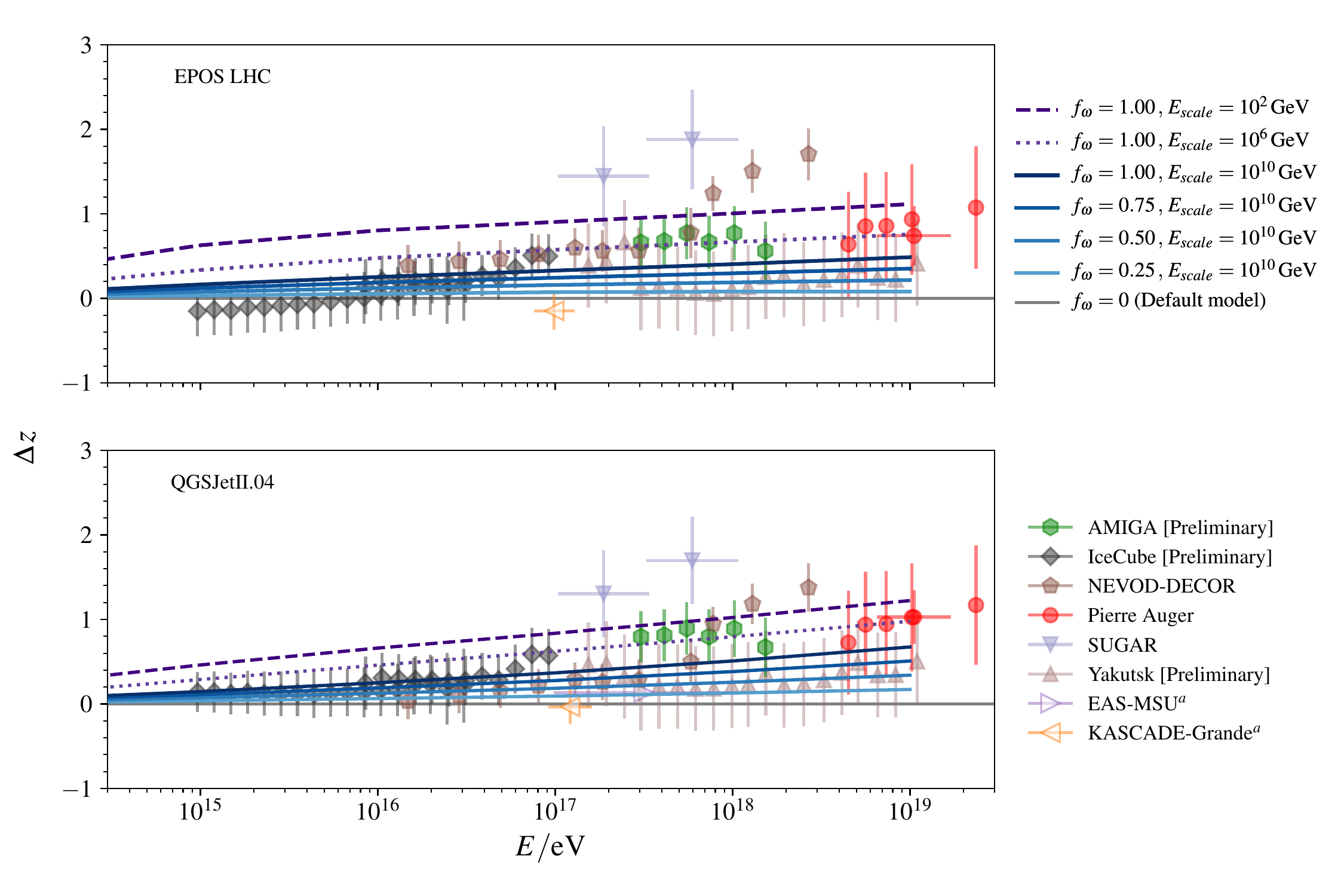}
  \caption{Evolution of the mass corrected $z$-factor, $\Delta z=z-z_{\mathrm mass}$, as a
    function of the primary energy. The data are taken from
    Ref.~\cite{Dembinski:2019uta} and references therein. Overlayed
    are predictions obtained from changing the scale $f_\omega$ (solid
    lines) and $\EsubText{scale}$ (dashed and dotted lines) obtained with
    \EPOSLHC{} (top) and \QGSJETIId{} (bottom) air shower
    simulations.}
  \label{fig:z_factor}
\end{figure*}

As pointed out in~Ref.~\cite{Dembinski:2019uta}, for all models the data have a positive $\Delta z$ showing a significant logarithmic increase with the primary energy, indicating an increasing muon deficit in the simulations.
In Fig.~\ref{fig:z_factor} the effect of the different energy
evolution of $\fcore$ for \EPOSLHC{} and \QGSJETIId{} on $\Delta z$ are shown.
Here the new simulations are treated like data and the $z$ factor is calculated
using the original (quoted) models as a reference such that the new $\Delta z$ can be
compared to the data points directly. The positive $\Delta z$ of the lines
indicate a larger muon production when $\fcore$ increases and the positive
slopes mean that the slope of the muon production as a
function of the primary energy is larger when $\fcore$ increases.
By including a consistent
core-like hadronization, we thus reproduce the energy evolution as found in the data.
This is even possible for values $\fcore<1$.

The possibility to see the effect of a core hadronization (QGP or
similar more exotic phenomena) on air shower physics have already been
studied in the
literature~\cite{Soriano:2017bvs,AlvarezMuniz:2012dd,Farrar:2013sfa,Anchordoqui:2016oxy}. Changes
in the muon production because of a change of $R$ under either extreme
or exotic assumptions (which were not yet observed at the LHC) are
usually assumed. Furthermore, it was shown that the production of a
core only in very central, high-density, collisions is not sufficient
to significantly change the muon numbers in air shower
simulations~\cite{LaHurd:2017pmb}.

In contrast to the new results presented here, in those previous
studies the core-like production does not cover sufficient phase space
to change the muon production in air showers significantly. We
demonstrate that core-like effects potentially starting at much
smaller colliding systems, and at much lower center-of-mass energies
as studied here, have an important impact on muon production in air
showers.  There are various indications at the LHC in $\Pproton{}\Pproton{}$ and $\Pproton{}\mathrm{A}$
collisions that such a scenario is compatible with current
data~\cite{Khachatryan:2010gv,ALICE:2017jyt}, or even suggested by it,
at energy densities as reached by cosmic rays interacting with the
atmosphere~\cite{Anchordoqui:2019laz}. Studying LHC data at mid-rapdity it is found that 
for events with $\langle{\rm d}N_{\rm ch}/{\rm d}\eta\rangle_{|\eta|<0.5}\sim10$ (corresponding to typical proton-air
interactions) $\fcore$ is already $\approx50$--$75\%$.
Since our study is based on the
simple assumption that the full phase space has a modified $\pi^0$
ratio, it remains crucial for cosmic ray physics to conduct further
dedicated measurements at the LHC to better understand $\pi^0$
production relative to other particles. The phase space for the
formation of core-like effects is potentially significantly larger
than previously studied, and in particular may extend towards larger
rapidities.

\section{Testing core contributions via measurements of $R$ at the LHC}
\label{sec:LHC}

As previously outlined an enhanced contribution of core-like hadronization can
help to explain the data of the Pierre Auger Observatory. In the following we
discuss how this can be probed with accelerator data.

We mainly use \EPOSLHC{} as the baseline model to test
sensitivity towards a QGP-like state. As alternative model we use
\PYTHIA8{}~\cite{Sjostrand:2006za,Sjostrand:2014zea}, which provides
entirely different (non-QGP-like) physics concepts for collectivity. \EPOSLHC{} is a
general purpose event generator widely used in high energy physics,
and in particular also for heavy ion collisions. It includes the
description of a QGP-like behavior in high energy
collisions. \PYTHIA8{}, on the other hand, is the reference model in
high energy physics for proton-proton interactions. Both models
generate a distribution of colored strings from the collision of a
projectile and a target. Despite a very different underlying approach
for the string generation (pQCD factorization for \PYTHIA8{} and
parton-based Gribov-Regge theory~\cite{Drescher:2000ha} for
\EPOSLHC{}), the string distributions are not very different, because
they are strongly constrained by the data on particle
multiplicities. These strings can be hadronized directly in both
generators using the Lund string model~\cite{Andersson:1983ia} in
\PYTHIA8{}, or the area law~\cite{Drescher:2000ha} in \EPOSLHC{} --
both cases are strongly constrained by LEP data. At low energy
($\approx$10 to 100\,GeV) this is sufficient to successfully describe
proton-proton interactions with good accuracy. Nevertheless, it turns
out that at the LHC additional physics mechanisms are needed to
describe the observed particle correlations and abundances in the
final state. In \PYTHIA8{}, a modified color reconnection
approach~\cite{Ortiz:2013yxa,Bierlich:2015rha} or a ``string shoving''
mechanism~\cite{Bierlich:2017vhg} have been proposed to introduce
collective effects such as a modified hadronization or particle
correlations, similar to those obtained from a QGP. In \EPOSLHC{}, on
the other hand, the ``core-corona'' approach~\cite{Werner:2007bf}
is used as originally developed for heavy ion collisions. As already
explained, the core amounts to areas with high string/energy densities,
where strings are assumed to ``melt'' and produce matter that expands
hydrodynamically and then decays statistically, whereas the corona
represents particles from ordinary string fragmentation, which escape
from the dense regions. While in \EPOSThree{}~\cite{Werner:2013tya}
the hydrodynamic expansion is fully implemented and hadronization
occurs on a freeze-out hypersurface, in \EPOSLHC{} this expansion is
mimicked by parameterizing the flow at hadronization.  This has proven
to describe various collective observables
well~\cite{Pierog:2013ria}. Simulations of \EPOSLHC{} are readily
available via the \textsc{crmc} software~\cite{crmc}.
On generator level, we study particles with a lifetime c$\tau>1$\,cm,
which is consistent with most experimental detector designs.

\begin{figure}
  \includegraphics[width=\columnwidth]{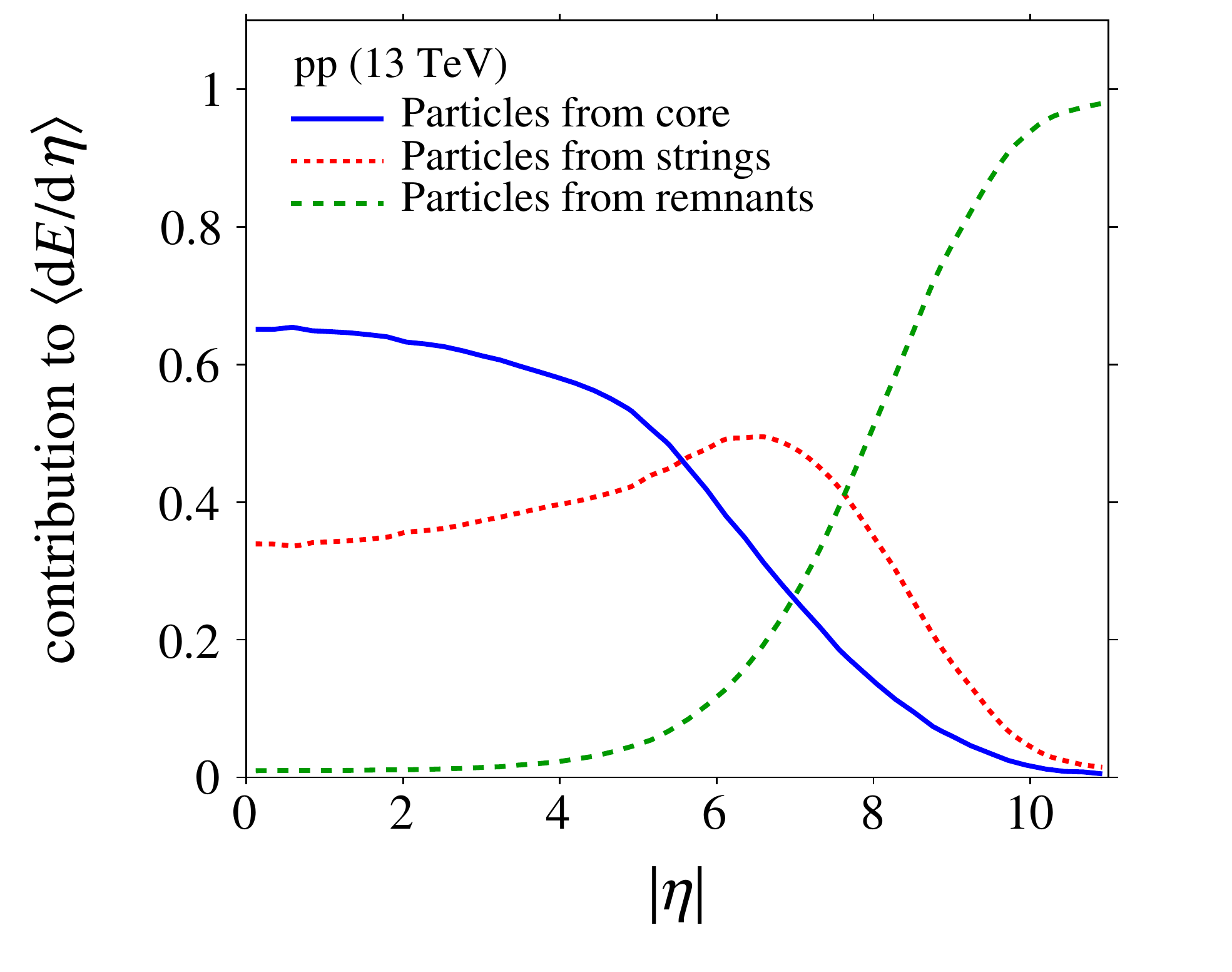}\\
  \includegraphics[width=\columnwidth]{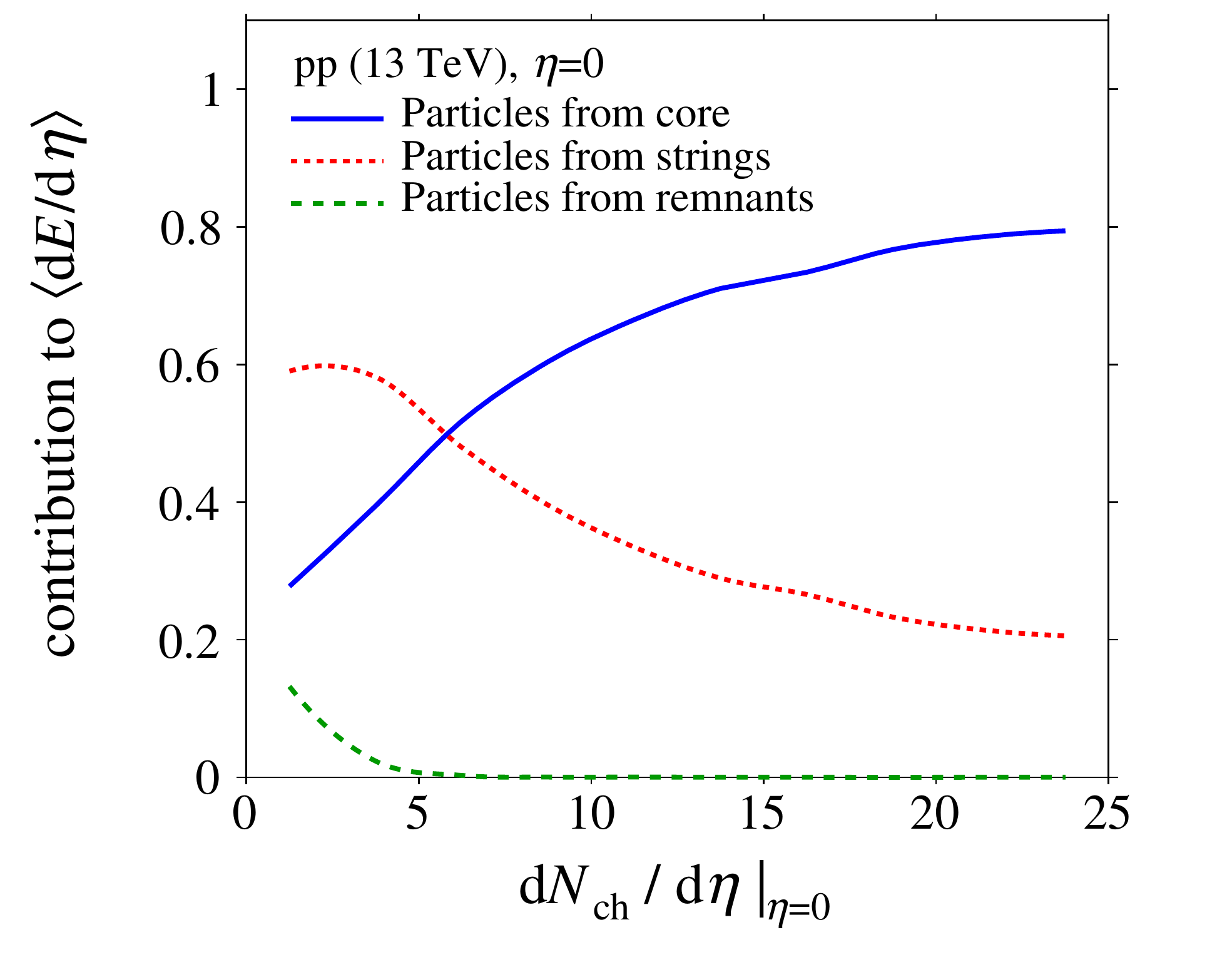}\\
  \includegraphics[width=\columnwidth]{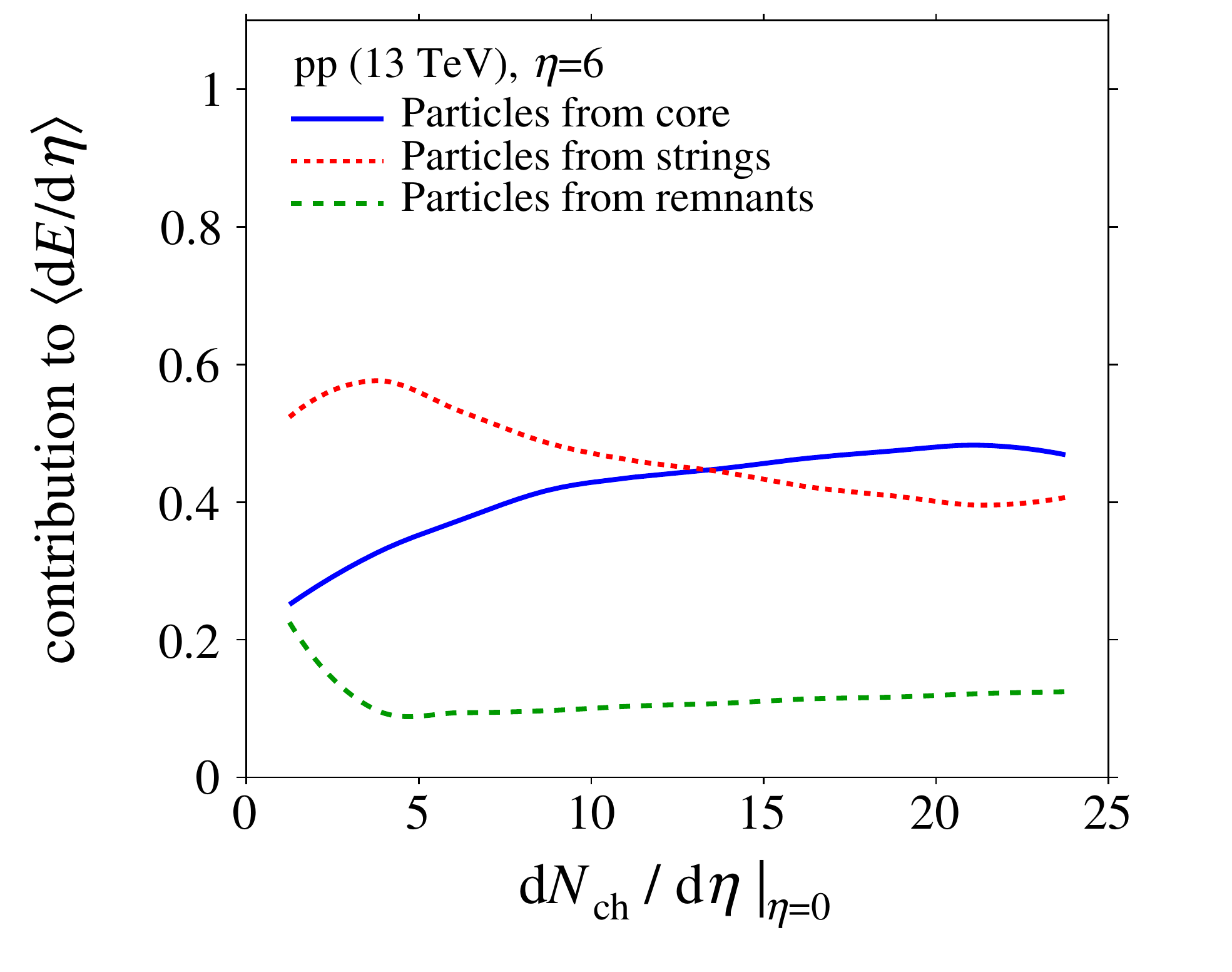}
  \caption{Fractional contribution of particles originating from
    different production mechanisms to the total energy density
    $\density{E}$ as predicted by \EPOSLHC{}. The top figure shows the
    contribution as function of $\abs{\eta}$; the middle (bottom)
    figure shows the contributions at $\eta=0$ ($\eta=6$) as a
    function of the charged particle density at $\eta=0$.}
\label{fig:EPOSEflowOrigin}
\end{figure}

In fact, in \EPOSLHC{} final-state particles originate from three different
production mechanisms: standard string fragmentation (corona),
statistical decay of a fluid (core), and the decay of the beam
remnants.  While experimentally the origin of the production mechanism
for a particle cannot be identified, individual
production mechanisms can still be studied since they predominantly
contribute to different regions of phase space.  This is demonstrated
in Fig.~\ref{fig:EPOSEflowOrigin}~(top), which shows the relative
contribution of these mechanisms to the total energy density
$\avg{\density{E}}$ for minimum bias proton-proton collisions at a
center-of-mass energy of 13\,TeV. Three regions can be identified: The
energy density at central pseudorapidities, $\abs{\eta}<5$, is
dominated by particles originating in the dense core of the
interaction, at intermediate rapidities, $5<\abs{\eta}<8$, it is
dominated by particles from string fragmentation, and at large
rapidities, $\abs{\eta}>8$, by the fragmentation of beam
remnants. Underlying differences in particle production, therefore,
lead to varying observables as a function of pseudorapidity.

A corresponding effect is also observed as a function of the central
charged particle multiplicity $\nch$. Final states with large particle
multiplicity are known to be an effective trigger for pronounced
statistical hadronization~\cite{Khachatryan:2010gv}. Therefore, at fixed pseudorapidity, the
influence of the core increases as a function of particle
multiplicity. This effect is expected to be most significant at
$\abs{\eta}\approx 0$ since the relative contribution of the core is
largest. This is illustrated in the middle panel of
Fig.~\ref{fig:EPOSEflowOrigin} for $\eta=0$ and in the bottom panel for
$\eta=6$. It can be seen that the contribution of the core to the
energy density at $\eta=0$ becomes dominant for $\Pproton{}\Pproton{}$ collisions with
more than $\approx7$ charged particles per unit of pseudorapidity,
while at $\eta=6$ this transition is shifted to a larger number.

\begin{figure}
\includegraphics[width=\columnwidth]{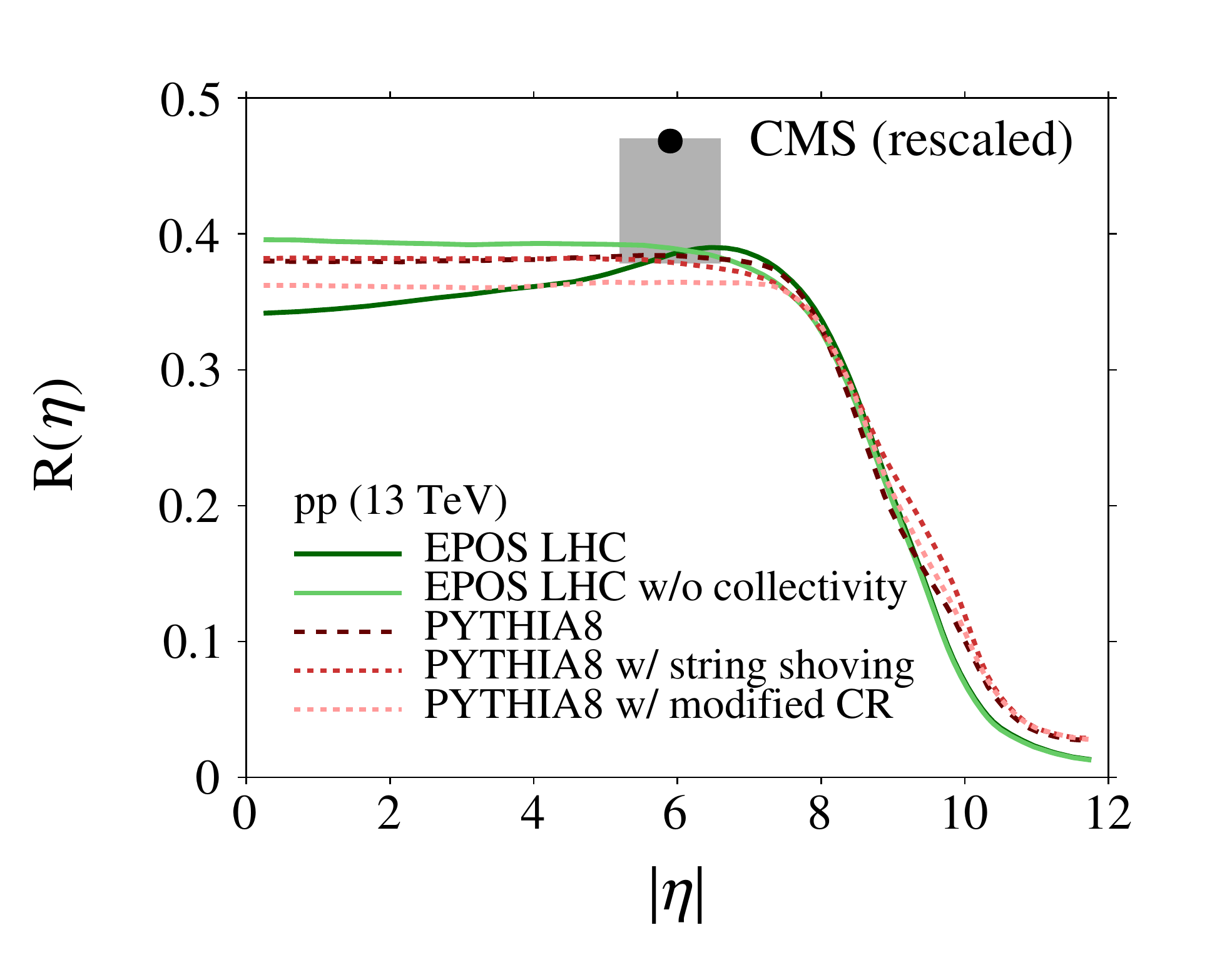}\\
\vspace{-0.3cm}
\includegraphics[width=\columnwidth]{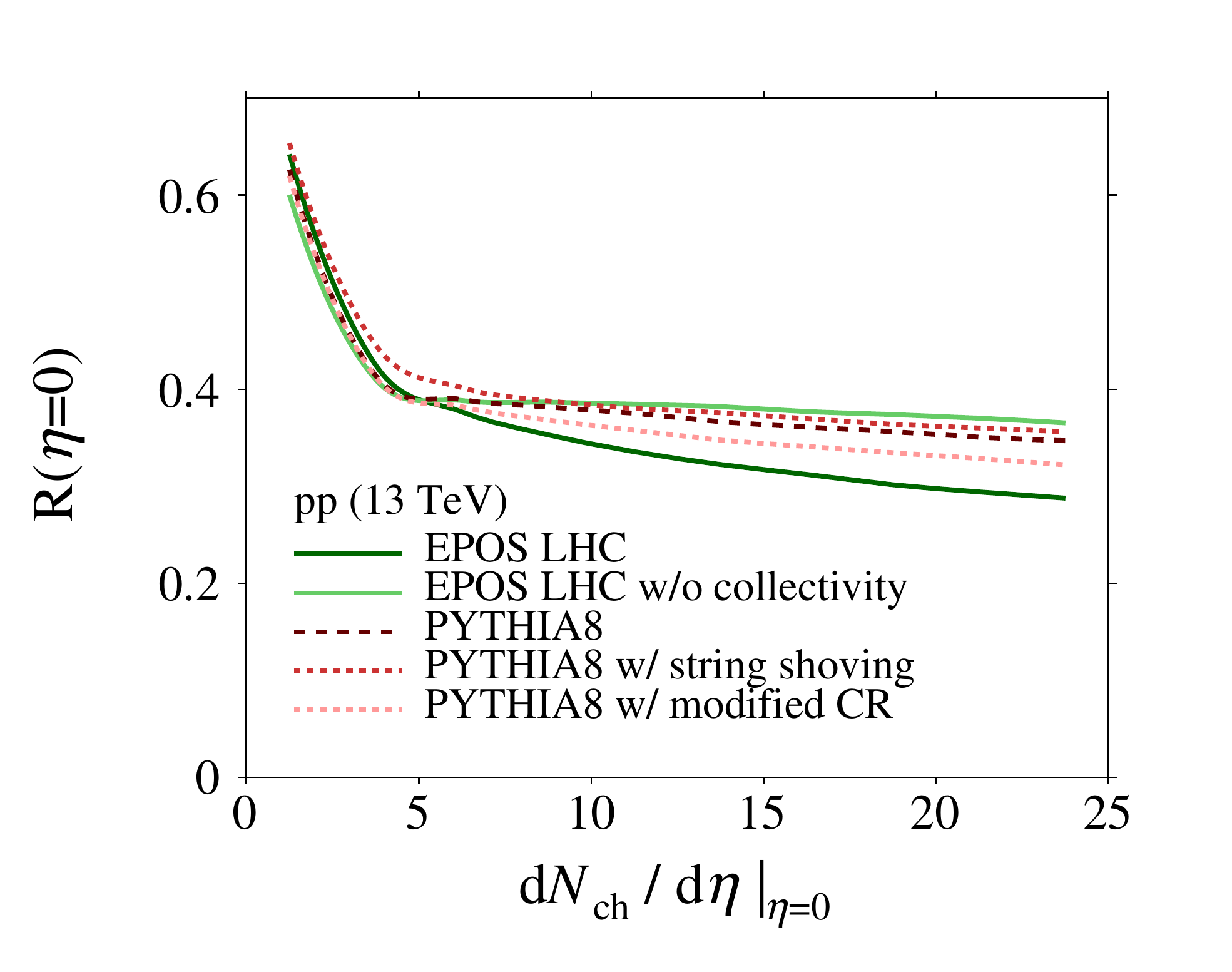}\\
\vspace{-0.3cm}
\includegraphics[width=\columnwidth]{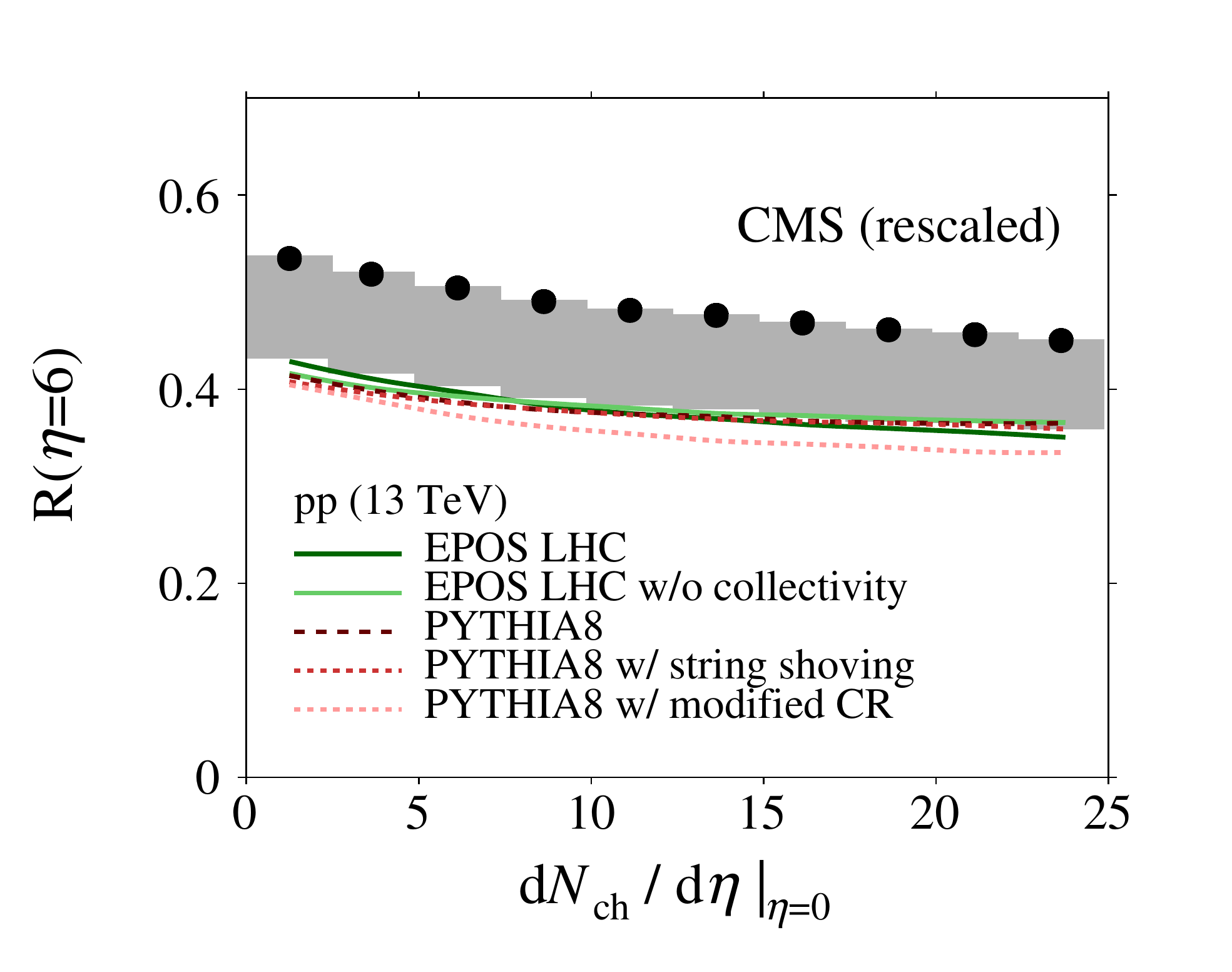}
\caption{Ratio of the average electromagnetic to hadronic energy
  densities $R$ simulated for proton-proton collisions at 13\,TeV with
  \EPOSLHC{} (solid lines) with and without hydrodynamical treatment
  of the dense core, as well as \PYTHIA8{} (dashed lines) in the
  default configuration, with string shoving and with modified color
  reconnection (CR). The top figure shows $R\left(\eta\right)$ as
  function of $\abs{\eta}$, the middle and bottom figure show $R$
  evaluated at $\eta=0$ and $\eta=6$ as a function of the central
  charged particle multiplicity. The asymmetric
  uncertainties of the CMS data are a feature of this measurement. }
\label{fig:EflowEmHad}
\end{figure}

Using \EPOSLHC{} we find that the fraction of secondary pions in the
dense core is reduced because many other more massive hadrons and
resonances are produced. This leads to a lower ratio of the
electromagnetic to hadronic energy density in particles produced from
the core. Accordingly, this effect can be seen in the
pseudorapidity-dependent ratio of the average electromagnetic to
hadronic energy density $R$ shown in top panel of
Fig.~\ref{fig:EflowEmHad}.  At $\abs{\eta}\approx 0$, the energy
density is dominated by the core and therefore the value of $R$ for
\EPOSLHC{} is as low as $0.34$. As the contribution of the core to the
total energy decreases with increasing pseudorapidity, also $R$
increases and reaches a value of $0.4$ at $\abs{\eta}\approx 7$ before
it decreases rapidly due to the very low electromagnetic contribution
in the beam remnants. In comparison, a flat ratio below
$\abs{\eta}\approx7$ is obtained when statistical hadronization is
disabled in \EPOSLHC{} (corona only). The data point shown in this figure
at $\eta\approx6$ is derived from Ref.~\cite{Sirunyan:2019rqy}, where
we have corrected the original values from detector level to generator
level using the Rivet routines provided by the CMS
Collaboration~\cite{Bierlich:2019rhm}. The shaded region
corresponds to the systematic uncertainties of the measurement. These
data are consistent with all models within the experimental
uncertainty; there is a slight tension with the $\PYTHIA8{}$
simulations using the modified colour reconnection
approach~\footnote{From \PYTHIA8{} manual: Option
  \texttt{ColourReconnection:mode=1} aka ``The new QCD based
  scheme''.}. Such data with smaller uncertainties, and measured over
a wide range of $\eta$ have the potential to differentiate between
some of the models. In particular, any slope observed in the region
$0<|\eta|\lesssim6$ would be a clear hint for a transition of several distinct 
hadronization mechanisms (i.e.~core-corona). 

The ratio of the electromagnetic to hadronic energy density $R$ at
$\eta=0$ is shown as a function of the central multiplicity
$\left.\density{N_{\mathrm{ch}}}\right|_{\eta=0}$ in the 
middle panel
of Fig.~\ref{fig:EflowEmHad}. It can be observed that $R$ drops down
to values of 0.3 when statistical hadronization is enabled in
\EPOSLHC{} while it reaches a constant plateau of 0.4 in the case of
disabled statistical hadronization, which is similar to the \PYTHIA8{} predictions. 
At $\eta=6$, it can be seen
in the bottom panel of Fig.~\ref{fig:EflowEmHad} how the different model
predictions compare to the available CMS data (also from Ref.~\cite{Sirunyan:2019rqy}).
However, these data are taken at $\eta\sim6$, were one can see from Fig.~\ref{fig:EflowEmHad} (top)
that the sensitivity to model differences is unfortunately close to minimal.
It would be a great way to study hadronization in hadron collisions by measuring this at LHC
in a much wider $\eta$ region.

We compare the simulations obtained with \EPOSLHC{} also to
predictions by \PYTHIA8{} in the standard minimum bias configuration
as well as with modified QCD-based color reconnection parameters
as presented in Ref.~\cite{Bierlich:2015rha}, and enabled string shoving
mechanisms. For the latter we use the example parameters provided
within \PYTHIA8{} version 8.235. We are aware that these settings are
untuned and results should be treated with care but first
observations, in particular about the characteristic shape of the
distributions, can be made. It is interesting that we do not find a
visible effect of the string shoving mechanism on the ratio of
electromagnetic to hadronic energy, $R$, compared to the default string
fragmentation (see Fig.~\ref{fig:EflowEmHad}). This is consistent with
the predictions of \EPOSLHC{} when statistical hadronization is disabled. These
findings can be understood since in all these cases particle
production is driven by QCD string fragmentation, which is well tuned
to LEP data. Thus, what is found here is a characteristic feature of
string fragmentation. If a microscopic collectivity model does not
modify particle production by string fragmentation, as it is the case
for string shoving, this has no impact on the observable $R$. With the
modified color reconnection on the other hand, a reduced value of $R$
is observed within $\abs{\eta}<7$ as well as a more prominent decrease
of $R$ at central rapidity as a function of
$\left.\density{N_{\mathrm{ch}}}\right|_{\eta=0}$. This is due to
enhanced baryon production as explained in
Ref.~\cite{Bierlich:2015rha}. Still, the decrease in $R$ is not as
strong as in \EPOSLHC{}. More importantly, all configurations of
\PYTHIA8{} exhibit a flat ratio as a function of pseudorapidity within
$\abs{\eta}<7$. The value of $R$ is a global feature of the
hadronization and independent of rapidity. No transition from a
statistical, to a string-dominated phase as in \EPOSLHC{} is observed.

\EPOSLHC{} was released after the first LHC data became available. At
that time, only average values and the evolution of the mean
transverse momentum as a function of the particle multiplicity were
known precisely. The increase of multi-strange baryon production with
particle multiplicity was a prediction of the model, but -- as shown
in Ref.~\cite{ALICE:2017jyt} -- was only qualitatively
correct. Effectively, the core is formed in \EPOSLHC{} only at larger
multiplicities compared to what is necessary to reproduce the data. Thus,
it is expected that the density needed to produce the core is
currently overestimated and, as a consequence, the effect on muon
production in air showers is significantly underestimated (not enough phase
space for core hadronization). 
It would be useful to have precise data on $R$ versus multiplicity to support (or reject) this hypothesis.

The study made with \EPOSLHC{} and \PYTHIA8{} is just an example of
what can be observed. A different model may have a different behavior,
but it has been clearly demonstrated that $R$ is sensitive to the type
of hadronization.  As a consequence, the observation of variations of
the ratio of electromagnetic to hadronic energies as function of
pseudorapidity or particle multiplicity is a strong test also of the nature
of collective effects in proton-proton collisions (or other systems).
Different implementations of statistical hadronization, QGP-like or
macroscopic, can be distinguished. The proposed measurements will
provide new constraints on the extension of the phase space in
which statistical hadronization occurs, complementary to established
measurements.

In any case, corresponding precision measurements of $R$ to 5\% at the
LHC seem feasible and could contribute significantly to a better
understanding of muon production in air showers as described in
section~\ref{sec:cr} in particular if the measurements could be done
with a light-ion beam such as oxygen~\cite{Citron:2018lsq}. Despite
the fact that calorimetric data are taken at various pseudorapidities
(central and forward calorimeters), such ratios are not commonly
published--with currently one notable exception~\cite{Sirunyan:2019rqy}. The reverse
argument also holds: in future huge-aperture air shower experiments,
the tail of the $\ln\nmu$-distribution could be used to indirectly
measure the slope of the energy distribution of neutral pions far
beyond the reach of the LHC~\cite{Cazon:2018bvs,Cazon:2018gww}.

\section{Summary}

We have demonstrated that the muon production in air shower
significantly depends on the ratio $R=\EsubText{em}/\EsubText{had}$,
where $\EsubText{em}$ is the sum of energy in secondary $\gamma$ (from $\pi^0$)
and $e^\pm$ while $\EsubText{had}$ is the sum of energy in
hadrons in individual hadron collisions.  We also showed that $R$ itself
depends on the hadronization mechanism. Thus, a
change or transition in these mechanisms can help to explain the
discrepancy between the observed number of muons in air showers by the
Pierre Auger Observatory and the predictions based on current hadronic
models. Since at the LHC, even in proton-proton interactions, one
observes a transition from a string-type to a statistical-type
hadronization at mid-rapidity, we used the particle ratios of the
statistical model at all pseudorapidities to show that such
hadronization scheme would in principle be sufficient to resolve the
observed difference between simulations and cosmic ray data. Experimental
measurements of $R$ at the LHC are currently compatible with this
possibility. On the other hand, extreme scenarios where full
statistical hadronization is reached at low energies
($\EsubText{lab}\sim\mathcal{O}(100\,$GeV$)$) are already excluded by the slope of the
energy-dependence of air shower muon data.

Furthermore, we discuss potential measurements of $R$ at LHC, e.g.~with
calorimeters, as a function of pseudorapidity $\eta$ or central charged
particle multiplicity $N_{\rm ch}$.  We show that this observable can
reveal properties of the nature of underlying fundamental particle
production mechanisms. In particular we show that it provides a new
handle to characterize mechanisms proposed for the explanation of
statistical hadronization in proton-proton collisions. It is potentially
possible to distinguish between quark-gluon-plasma-like (QGP-like) effects, as
first known from heavy ion collisions, from alternative, more
microscopic effects that do not require the formation of a QGP, see
e.g.\ Refs.~\cite{Bierlich:2015rha,Bierlich:2017vhg}.

Dedicated measurements at the LHC have now another opportunity to
study collectivity in proton-proton collision using this
observable. This will contribute to a better understanding of the
mechanisms of hadronization in hadron collisions, and collectivity in
proton-proton collisions or other light system. Measuring $R$ at the
LHC potentially has a significant impact on resolving the current
mystery of muon production in cosmic ray induced extensive air
showers. Thus, at last, one aspect to resolve the cosmic ray muon
mystery is a better understanding of statistical hadronization in small
collision systems.

\bibliography{main}

\end{document}